# A NEW RADIAL BASIS FUNCTION APPROXIMATION WITH REPRODUCTION

Zuzana Majdisova
*Faculty of Applied Sciences, University of West Bohemia*
*Univerzitni 8, CZ 30614 Plzen, Czech Republic*

Vaclav Skala
*Faculty of Applied Sciences, University of West Bohemia*
*Univerzitni 8, CZ 30614 Plzen, Czech Republic*

**ABSTRACT**

Approximation of scattered geometric data is often a task in many engineering problems. The Radial Basis Function (RBF) approximation is appropriate for large scattered (unordered) datasets in $d$-dimensional space. This method is useful for a higher dimension $d \geq 2$, because the other methods require a conversion of a scattered dataset to a semi-regular mesh using some tessellation techniques, which is computationally expensive. The RBF approximation is non-separable, as it is based on a distance of two points. It leads to a solution of overdetermined Linear System of Equations (LSE).

In this paper a new RBF approximation method is derived and presented. The presented approach is applicable for $d$-dimensional cases in general.

## 1. INTRODUCTION

Radial Basis Functions (RBFs) are widely used across many fields solving technical and non-technical problems. The RBF method was originally introduced by [Hardy, R.L., 1971] and it is an effective tool for solving partial differential equations in engineering and sciences. Moreover, RBF applications can be found in neural networks, fuzzy systems, pattern recognition, data visualization, medical applications, surface reconstruction [Carr, J.C. et al, 2001], [Turk, G. and O'Brien, J.F., 2002], [Pan, R. and Skala, V., 2011a], [Pan, R. and Skala, V., 2011b], [Skala, V. et al, 2013], [Skala, V. et al, 2014], reconstruction of corrupted images [Uhlir, K. and Skala, V., 2005], [Zapletal, J. et al, 2009], etc. The RBF approximation technique is really meshless and is based on collocation in a set of scattered nodes. This method is independent with respect to the dimension of the space. The computational cost of RBF approximation increases nonlinearly with the number of points in the given dataset and linearly with the dimensionality of data.

There are two main groups of basis functions: global RBFs (e.g. [Duchon, J., 1977], [Schagen, I.P, 1979]) and Compactly Supported RBFs (CS-RBFs) [Wendland, H., 2006]. Fitting scattered data with CS-RBFs leads to a simpler and faster computation, because the system of linear equations has a sparse matrix. However, approximation using CS-RBFs is sensitive to the density of approximated scattered data and to the choice of a "shape" parameter. Global RBFs lead to a linear system of equations with a dense matrix and their usage is based on sophisticated techniques such as the fast multipole method [Darve, E., 2000]. Global RBFs are useful in repairing incomplete datasets and they are significantly less sensitive to the density of approximated data.

## 2. ORIGINAL APPROACH

The original approach of RBF approximation with linear reproduction was introduced by [Fasshauer, G.E., 2007] (Chapter 19.4). Let us briefly summarize the properties of this approach in this section.

The goal of this approach is to approximate a given dataset of $N$ points by a function:

$$f(x) = \sum_{j=1}^{M} c_j \phi(\|x - \xi_j\|) + P_1(x), \tag{1}$$

where the approximating function $f(x)$ is represented as a sum of $M$ RBFs, each associated with a different reference point $\xi_j$, and weighted by an appropriate coefficient $c_j$, and $P_1(x) = a^T x + a_0$ is a linear polynomial. This linear polynomial should theoretically solve problems with stability and solvability. Now, it is necessary to determine the vector of weights $c = (c_1, \ldots, c_M)^T$ and coefficients of the linear polynomial. This is achieved by solving an overdetermined linear system of equations (LSE):

$$h_i = f(x_i) = \sum_{j=1}^{M} c_j \phi(\|x_i - \xi_j\|) + P_1(x_i) = \sum_{j=1}^{M} c_j \phi_{ij} + P_1(x_i), \quad i = 1, \ldots, N, \tag{2}$$

where $x_i$ is point from the given dataset and is associated with scalar value $h_i$. Moreover, additional conditions are applied:

$$\sum_{i=1}^{M} c_i = 0, \quad \sum_{i=1}^{M} c_i \xi_i = \mathbf{0}. \tag{3}$$

It can be seen that for $d$-dimensional space a linear system of $(N + d + 1)$ equations in $(M + d + 1)$ variables has to be solved, where $N$ is the number of points in the given dataset, $M$ is the number of reference points and $d$ is the dimensionality of the data.

For $d = 2$, vectors $x_i, \xi_j$ and $a$ are given as $x_i = (x_i, y_i)^T$, $\xi_j = (\xi_j, \eta_j)^T$ and $a = (a_x, a_y)^T$. Thus, for $E^2$ and the given dataset we can write this LSE in the following matrix form:

$$\begin{pmatrix} A & P \\ \Xi & 0 \end{pmatrix} \begin{pmatrix} c \\ a \\ a_0 \end{pmatrix} = \begin{pmatrix} h \\ 0 \end{pmatrix} \tag{4}$$

This system is overdetermined ($M \ll N$) and can be solved by the least squares method as:

$$\begin{pmatrix} A^T A + \Xi^T \Xi & A^T P \\ P^T A & P^T P \end{pmatrix} \begin{pmatrix} c \\ a \\ a_0 \end{pmatrix} = \begin{pmatrix} A^T h \\ P^T h \end{pmatrix} \tag{5}$$

where

$$A^T A + \Xi^T \Xi = \begin{pmatrix} \sum_{i=1}^{N} \phi_{i1} \phi_{i1} + \xi_1^2 + \eta_1^2 + 1 & \cdots & \sum_{i=1}^{N} \phi_{i1} \phi_{iM} + \xi_1 \xi_M + \eta_1 \eta_M + 1 \\ \vdots & \ddots & \vdots \\ \sum_{i=1}^{N} \phi_{iM} \phi_{i1} + \xi_M \xi_1 + \eta_M \eta_1 + 1 & \cdots & \sum_{i=1}^{N} \phi_{iM} \phi_{iM} + \xi_M^2 + \eta_M^2 + 1 \end{pmatrix},$$

$$P^T A = (A^T P)^T = \begin{pmatrix} \sum_{i=1}^{N} x_i \phi_{i1} & \cdots & \sum_{i=1}^{N} x_i \phi_{iM} \\ \sum_{i=1}^{N} y_i \phi_{i1} & \cdots & \sum_{i=1}^{N} y_i \phi_{iM} \\ \sum_{i=1}^{N} \phi_{i1} & \cdots & \sum_{i=1}^{N} \phi_{iM} \end{pmatrix}, \quad P^T P = \begin{pmatrix} \sum_{i=1}^{N} x_i^2 & \sum_{i=1}^{N} x_i y_i & \sum_{i=1}^{N} x_i \\ \sum_{i=1}^{N} y_i x_i & \sum_{i=1}^{N} y_i^2 & \sum_{i=1}^{N} y_i \\ \sum_{i=1}^{N} x_i & \sum_{i=1}^{N} y_i & \sum_{i=1}^{N} 1 \end{pmatrix},$$

$$A^T h = \left( \sum_{i=1}^{N} \phi_{i1} h_i \quad \cdots \quad \sum_{i=1}^{N} \phi_{iM} h_i \right)^T, \quad P^T h = \left( \sum_{i=1}^{N} x_i h_i \quad \sum_{i=1}^{N} y_i h_i \quad \sum_{i=1}^{N} h_i \right)^T.$$

It should be noted that additional conditions (3) introduce inconsistency to the least squares method. Specifically, the inconsistency is caused by adding the term $\Xi^T \Xi$ to $A^T A$. Therefore, the described RBF approximation with linear reproduction is inconveniently formulated, as it mixes variables which have a different physical meaning. Thus, another approach is proposed in the following section.

## 3. PROPOSED APPROACH

Let us consider that we have an unordered dataset $\{x_i\}_1^N$ in $E^2$. However, note that this approach is generally applicable for $d$-dimensional space. Further, each point $x_i$ from the dataset is associated with vector $h_i \in E^p$ of given values, where $p$ is the dimension of the vector, or a scalar value $h_i \in E^1$. For an explanation of the RBF approximation, let us consider the case in which each point $x_i$ is associated with a scalar value $h_i$, e.g. a 2 ½D surface. Let us introduce a set of new reference points $\{\xi_j\}_1^M$, see Figure 1.

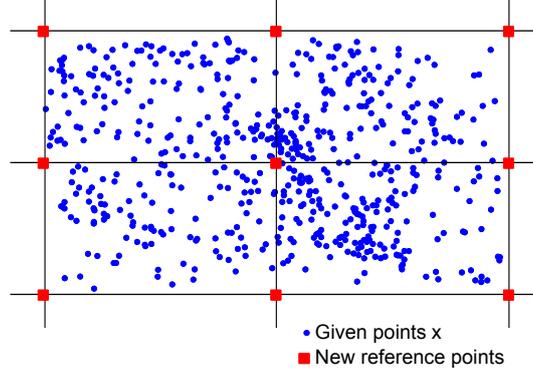

Figure 1: RBF approximation and reduction of points.

It should be noted that these reference points may not necessarily be in a uniform grid. It is appropriate that their placements reflect the given surface behavior (e.g. the terrain profile, etc.) as well as possible. The number of added reference points $\xi_j$ is $M$, where $M \ll N$. The RBF approximation is based on computing the distance of the given point $x_i$ of the given dataset and the reference point $\xi_j$ of the new reference points.

The approximated value can be expressed as:

$$f(x) = \sum_{j=1}^{M} c_j \, \phi(\|x - \xi_j\|) \; + P_1(x), \tag{6}$$

where the approximating function $f(x)$ is represented as a sum of $M$ RBFs, each associated with a different reference point $\xi_j$, and weighted by an appropriate coefficient $c_j$, and $P_1(x) = a^T x + a_0$ is a linear polynomial. This linear polynomial should theoretically solve problems with stability and solvability.

It can be seen that for $E^2$ and the given dataset we get the following overdetermined LSE:

$$Ac + Pk = h, \tag{7}$$

where $A_{ij} = \phi(\|x_i - \xi_j\|)$ is the entry of the matrix in the $i$-th row and $j$-th column, $c = (c_1, \ldots, c_M)^T$ is the vector of weights, $P_i = (x_i^T, 1)$ is the vector, $k = (a^T, a_0)^T$ is the vector of coefficients for the linear polynomial and $h = (h_1, \ldots, h_N)^T$ is the vector of values in the given points.

The error is then defined as:

$$R = \|Ac + Pk - h\|, \tag{8}$$

then

$$R^2 = (Ac + Pk - h)^T(Ac + Pk - h). \tag{9}$$

Our goal is to minimize the square of error, i.e. to find the minimum of $R^2$ (9). This minimum is obtained by differentiating equation (9) with respect to $c$ and $k$ and finding the zeros of those derivatives. This leads to equations:

$$\frac{\partial R^2}{\partial c} = 2(A^T A c + A^T P k - A^T h) = 0,$$

$$\frac{\partial R^2}{\partial k} = 2(P^T A c + P^T P k - P^T h) = 0, \tag{10}$$

which leads to a system of linear equations:

$$\begin{pmatrix} A^T A & A^T P \\ P^T A & P^T P \end{pmatrix} \begin{pmatrix} c \\ k \end{pmatrix} = \begin{pmatrix} A^T h \\ P^T h \end{pmatrix}, \tag{11}$$

i.e.
$$B\lambda = f. \tag{12}$$

The matrix $B$ is a $(M+3) \times (M+3)$ symmetric positively semidefinite matrix. Equation (11) can be expressed in the form:

$$\begin{pmatrix} \sum_{i=1}^{N}\phi_{i1}\phi_{i1} & \cdots & \sum_{i=1}^{N}\phi_{i1}\phi_{iM} & \sum_{i=1}^{N}\phi_{i1}x_i & \sum_{i=1}^{N}\phi_{i1}y_i & \sum_{i=1}^{N}\phi_{i1} \\ \vdots & \ddots & \vdots & \vdots & \vdots & \vdots \\ \sum_{i=1}^{N}\phi_{iM}\phi_{i1} & \cdots & \sum_{i=1}^{N}\phi_{iM}\phi_{iM} & \sum_{i=1}^{N}\phi_{iM}x_i & \sum_{i=1}^{N}\phi_{iM}y_i & \sum_{i=1}^{N}\phi_{iM} \\ \sum_{i=1}^{N}x_i\phi_{i1} & \cdots & \sum_{i=1}^{N}x_i\phi_{iM} & \sum_{i=1}^{N}x_i^2 & \sum_{i=1}^{N}x_iy_i & \sum_{i=1}^{N}x_i \\ \sum_{i=1}^{N}y_i\phi_{i1} & \cdots & \sum_{i=1}^{N}y_i\phi_{iM} & \sum_{i=1}^{N}y_ix_i & \sum_{i=1}^{N}y_i^2 & \sum_{i=1}^{N}y_i \\ \sum_{i=1}^{N}\phi_{i1} & \cdots & \sum_{i=1}^{N}\phi_{iM} & \sum_{i=1}^{N}x_i & \sum_{i=1}^{N}y_i & \sum_{i=1}^{N}1 \end{pmatrix} \begin{pmatrix} c_1 \\ \vdots \\ c_M \\ a_x \\ a_y \\ a_0 \end{pmatrix} = \begin{pmatrix} \sum_{i=1}^{N}\phi_{i1}h_i \\ \vdots \\ \sum_{i=1}^{N}\phi_{iM}h_i \\ \sum_{i=1}^{N}x_ih_i \\ \sum_{i=1}^{N}y_ih_i \\ \sum_{i=1}^{N}h_i \end{pmatrix}. \tag{13}$$

where $\phi_{ij} = \phi(\|x_i - \xi_j\|)$, point $x_i = (x_i, y_i)^T$ and vector $a = (a_x, a_y)^T$. It can be seen that this approach eliminates the inconsistency introduced in Section 2.

## 4. EXPERIMENTAL RESULTS

Both presented approaches of the RBF approximation have been compared for a dataset with a Halton distribution of points [Fasshauer, G.E., 2007] (Appendix A.1). Moreover, each point from this dataset is associated with a function value at this point. For this purpose, different functions have been used for experiments. Results for two such functions are presented here. The first is a $2D$ sinc function, see Figure 2 (left), and the second is Franke's function, see Figure 2 (right).

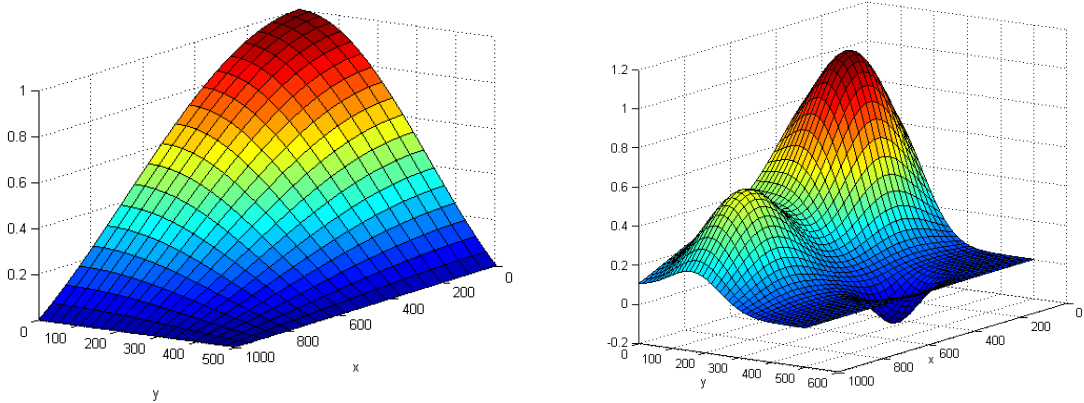

Figure 2: $2D$ sinc function defined as $\text{sinc}\left(\frac{\pi x}{1000}\right)\text{sinc}\left(\frac{\pi y}{500}\right)$, whose domain is restricted to $[0,1000] \times [0,500]$ (left) and Franke's function (right).

In addition, three different global radial basis functions with shape parameter $\alpha$, see Table 1, have been used for testing. Also different sets of reference points have been used for experiments.

Table 1. Used global RBFs.

| RBF | $\phi(r)$ |
|---|---|
| **Gauss function** | $e^{-(\alpha r)^2}$ |
| **Inverse Quadric (IQ)** | $\dfrac{1}{1+(\alpha r)^2}$ |
| **Thin-Plate Spline (TPS)** | $(\alpha r)^2 \log(\alpha r)$ |

These sets of reference points have different types of distributions. The presented types of distribution are the Halton distribution [Fasshauer, G.E., 2007] (Appendix A.1), see Figure 3 (left), an epsilon distribution, which is based on a random drift of points on a regular grid, see Figure 3 (right), and points on a regular grid.

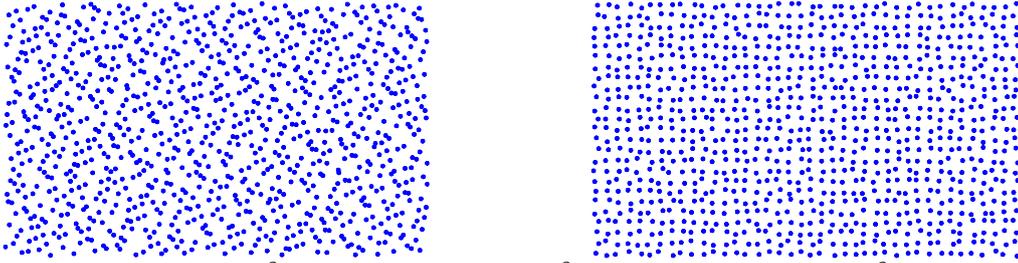

Figure 3: Halton points in $E^2$ (left) and epsilon points in $E^2$ (right). Number of points is $10^3$ in both cases.

## 4.1 Examples of RBF Approximation Results

An example of RBF approximation of 1089 Halton data points sampled from a $2D$ sinc function, for a Halton set of reference points which consists of 81 points, using both approaches is shown in Figure 4. The graphs are false-colored according to the magnitude of the error.

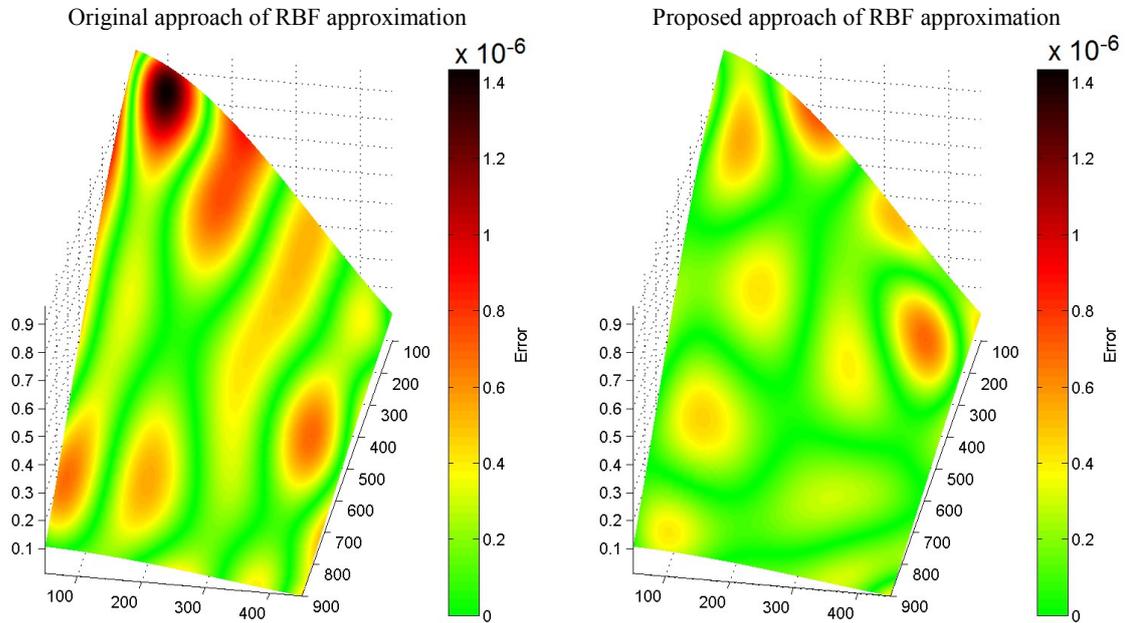

Figure 4: Approximation of 1089 data points sampled from a $2D$ sinc function, i.e. $\text{sinc}\left(\dfrac{\pi x}{1000}\right) \text{sinc}\left(\dfrac{\pi y}{500}\right)$, where $(x, y) \in [0,1000] \times [0,500]$, with 81 Halton-spaced Gaussian functions with $\alpha = 0.001$, false-colored by magnitude of error.

A further example of RBF approximation of 4225 Halton data points sampled from a Franke's function and for a set of reference points which consists of 289 points on a regular grid, using both approaches is shown in Figure 5. The graphs are again false-colored by magnitude of error.

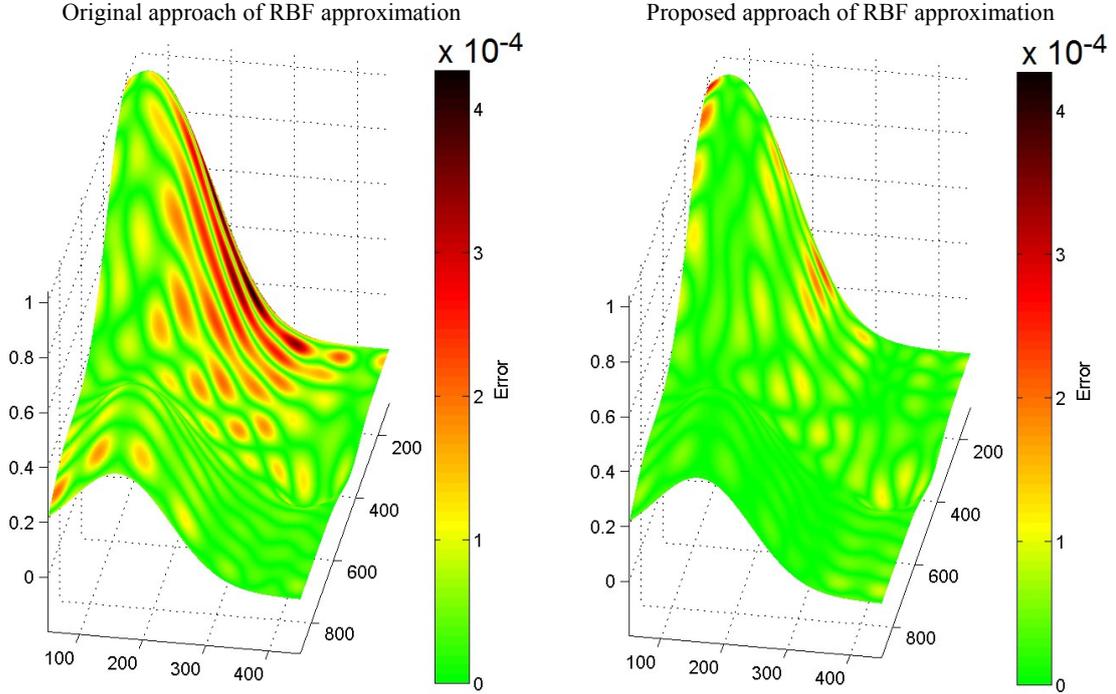

Figure 5: Approximation of 4225 data points sampled from a Franke's function with 289 regularly spaced IQ with $\alpha = 0.005$, false-colored by magnitude of error.

It can be seen that the original RBF approximation with a linear reproduction returns a worse result in terms of the error in comparison with the proposed RBF approximation with a linear reproduction. Moreover, we can see from Figure 4 and Figure 5 that for the presented cases **the maximum magnitude of error for the original approach is approximately two times greater than the maximum magnitude of error for the proposed approach**.

There remains the question of how the RBF approximation depends on the shape parameter $\alpha$ selection. Many papers have been published about choosing optimal shape parameter $\alpha$, e.g. [Franke, R., 1982], [Rippa, S., 1999], [Fasshauer, G.E. and Zhang, J.G., 2007], [Scheuerer, M., 2011]. In the following section, a comparison depending on the choice of shape parameter $\alpha$ is performed.

## 4.2 Comparison of Methods

In this section, the original approach and the proposed approach, which were presented in Section 2 and Section 3, are compared. Figure 6 presents the ratio of mean error of the original RBF approximation with the linear reproduction to the mean error of the proposed RBF approximation with the linear reproduction, i.e.:

$$ratio = \frac{mean\ error_{original}}{mean\ error_{proposed}}, \quad (14)$$

for a dataset which consists of 1089 Halton points in the range $[0,1000] \times [0,500]$, sampled from a $2D$ sinc function. The set of reference points contains 81 points with different behavior of the distribution, and for different global RBFs. Graphs in Figure 6 represent the experimentally obtained ratio according to the shape parameter $\alpha$ of the used RBFs.

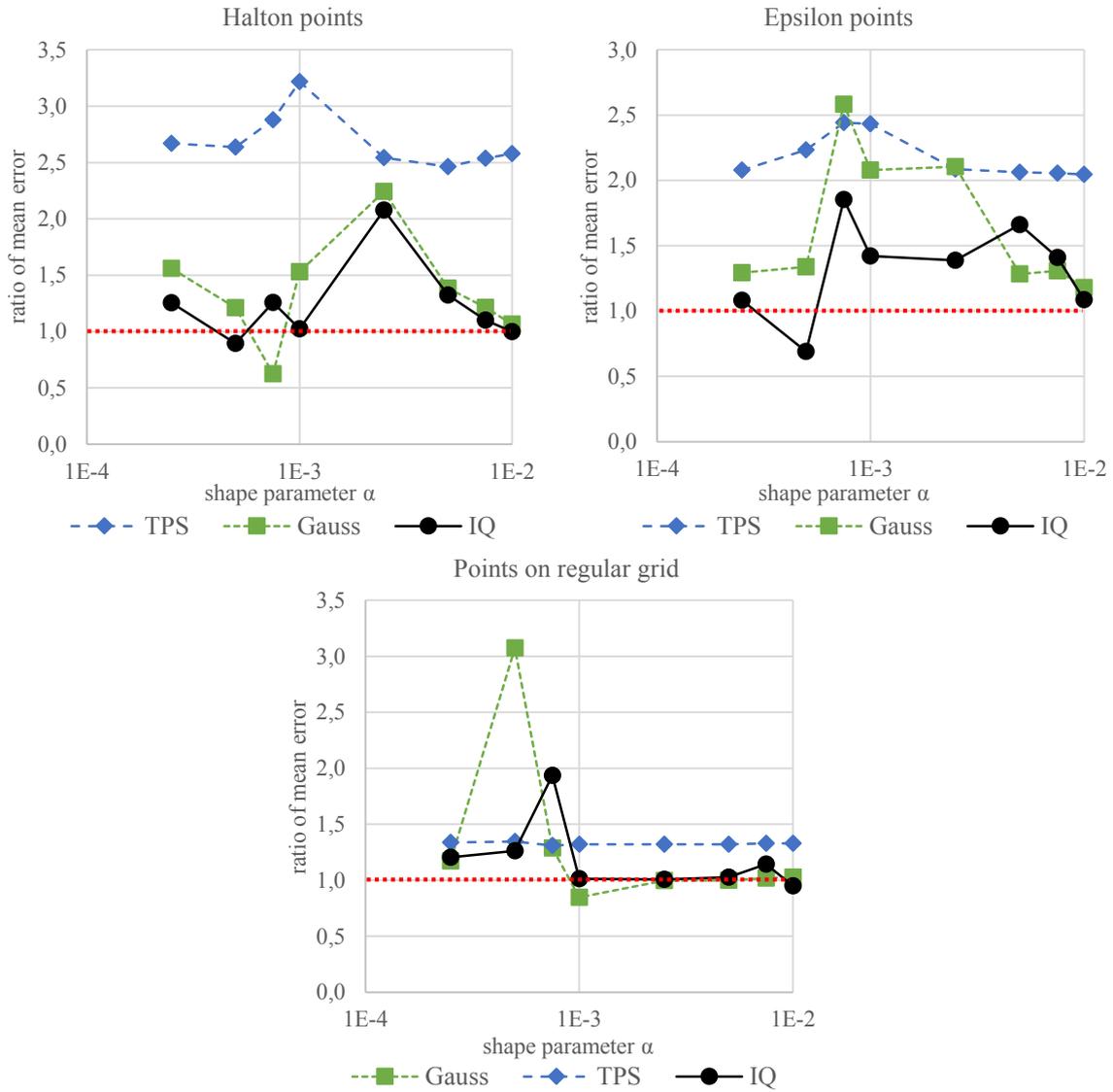

Figure 6: The ratio of mean error of the original approach to the mean error of the proposed approach of RBF approximation of 1089 data points sampled from a $2D$ sinc function with 81 reference points for different RBFs and different shape parameters. The used sets of reference points are: Halton points (top left), Epsilon points (top right) and points on a regular grid (bottom).

We can see that for the TPS, the mean errors of the proposed approach are significantly smaller than those of the original approach (ratio is greater than one). Furthermore, this ratio is not significantly different for the different shape parameters $\alpha$. For the Gaussian function and epsilon reference points, the proposed RBF approximation gives better results than the original approach in terms of the mean error. In the remaining cases, with five exceptions, the proposed approach is also better.

The experiments prove that the proposed approach to RBF approximation is correct and gives better and more stable results than the original approach [Fasshauer, G.E., 2007].

## 5. CONCLUSION

This paper presents a new formulation for RBF approximation with a linear reproduction. The proposed approach eliminates inconsistency, which occurs in the original RBF approximation with a linear reproduction. This inconsistency is caused by adding additional conditions to the polynomial part. The experiments made prove that the proposed approach gives significantly better results than the original method in terms of accuracy. The presented approach is easily extendable for general polynomial reproduction and for higher dimensionality.

In future work, application of the proposed approach is to be performed on large real datasets and the performance can be further measured.

## ACKNOWLEDGEMENT

The authors would like to thank their colleagues at the University of West Bohemia, Plzen, for their discussions and suggestions, and also anonymous reviewers for the valuable comments and suggestions they provided. The research was supported by MSMT CR projects LH12181 and SGS 2016-013.

## REFERENCES

Carr, J.C. et al, 2001. Reconstruction and representation of 3D objects with radial basis functions. *Proceedings of the 28th Annual Conference on Computer Graphics and Interactive Techniques, SIGGRAPH 2001*, Los Angeles, California, USA, pp. 67-76.

Darve, E., 2000. The Fast Multipole Method: Numerical Implementation. *In Journal of Computational Physics*, Vol. 160, No. 1, pp. 195-240.

Duchon, J., 1977. Splines Minimizing Rotation-invariant Semi-norms in Sobolev Spaces. *In Constructive theory of functions of several variables*, pp. 85-100.

Fasshauer, G.E., 2007. *Meshfree Approximation Methods with MATLAB*. World Scientific Publishing Co., River Edge, NJ, USA.

Fasshauer, G.E. and Zhang, J.G., 2007. On choosing optimal shape parameters for RBF approximation. *In Numeric Algorithms*, Vol. 45, pp. 345-368.

Franke, R., 1982. Scattered data interpolation: tests of some methods. *In Mathematical Computing*, Vol. 38, pp. 181-200.

Hardy, R.L., 1971. Multiquadratic Equations of Topography and Other Irregular Surfaces. *In Journal of Geophysical Research*, Vol. 76, No. 8, pp. 1905-1915.

Pan, R. and Skala, V., 2011a. Continuous Global Optimization in Surface Reconstruction from an Oriented Point Cloud. *In Computer-Aided Design*, Vol. 43, No. 8, pp. 896-901.

Pan, R. and Skala, V., 2011b. A Two-level Approach to Implicit Surface Modeling with Compactly Supported Radial Basis Functions. *In Eng. Comput. (Lond.)*, Vol. 27, No. 3, pp. 299-307.

Rippa, S., 1999. An algorithm for selecting a good value for the parameter c in radial basis function interpolation. *In Adv. Comput. Math.*, Vol. 11, pp. 193-210.

Schagen, I.P, 1979. Interpolation in Two Dimensions - a New Technique. *In IMA Journal of Applied Mathematics*, Vol. 23, No. 1, pp. 53-59.

Scheuerer, M., 2011. An alternative procedure for selecting a good value for the parameter c in RBF-interpolation. *In Adv. Comput. Math.*, Vol. 34, pp. 105-126.

Skala, V. et al, 2013. Simple 3D Surface Reconstruction Using Flatbed Scanner and 3D Print. *SIGGRAPH Asia 2013, Poster Proceedings.* Hong Kong, China, p. 7.

Skala, V. et al, 2014. Making 3D Replicas Using a Flatbed Scanner and a 3D Printer. *Computational Science and Its Applications - ICCSA 2014 - 14th International Conference, Proceedings, Part VI.* Guimarães, Portugal, pp. 76-86.

Turk, G. and O'Brien, J.F., 2002. Modelling with implicit surfaces that interpolate. *In ACM Trans. Graph.*, Vol. 21, No. 4, pp. 855-873.

Uhlir, K. and Skala, V., 2005. Reconstruction of Damaged Images Using Radial Basis Functions. *Proceedings of EUSIPCO.* Antalya, Turkey, p. 160.

Wendland, H., 2006. Computational Aspects of Radial Basis Function Approximation. *In Studies in Computational Mathematics*, Vol. 12, pp. 231-256.

Zapletal, J. et al, 2009. RBF-based Image Restoration Utilizing Auxiliary Points. *Proceedings of the 2009 Computer Graphics International Conference.* Victoria, British Columbia, Canada, pp. 39-43.